\documentclass[aps,prb,superscriptaddress,twocolumn]{revtex4}%
\usepackage{bm}
\usepackage{psfig}%
\usepackage{amsmath}%
\usepackage{amsfonts}%
\usepackage{amssymb}%
\usepackage{graphicx}
\usepackage{enumerate}
\begin{document}

\newcommand{\ac}{{\tt a}}
\newcommand{\hr}{Hit\&Run }
\newcommand{\var}{\mathtt{Var}}
\newcommand{\eff}{\epsilon^*} 
\newcommand{\la}{\langle} 
\newcommand{\ra}{\rangle} 
\newcommand{\vol}{\Omega}
\newcommand{\ferm}{\mathcal{F}}
\newcommand{\fstar}{F_{w}}
\newcommand{\ferma}{\la \mathcal{F}_c \ra_{f}}
\newcommand{\fermac}{\la \mathcal{F}_c \ra_{h}}

\title{Determination of the chemical potential using energy-biased sampling}
\author{R. ~Delgado-Buscalioni}
\email[]{r.delgado-buscalioni@ucl.ac.uk}
\affiliation{Depto. Ciencias y T\'ecnicas Fisicoqu\'{\i}micas, Facultad de Ciencias, UNED, Paseo Senda del Rey 9,  Madrid 28040, Spain.}
\author{G. De Fabritiis}
\email[]{g.defabritiis@ucl.ac.uk}
\affiliation{Centre for Computational Science, Department of Chemistry, University College
London, 20 Gordon Street, WC1H 0AJ London, U.K.}
\author{P. V. Coveney}
\email[]{p.v.coveney@ucl.ac.uk}
\affiliation{Centre for Computational Science, Department of Chemistry, University College
London, 20 Gordon Street, WC1H 0AJ London, U.K.}
\date{\today}

\begin{abstract}
An energy-biased method to evaluate ensemble averages requiring
test-particle insertion is presented. The method is based on biasing
the sampling within the subdomains of the test-particle configurational space 
with energies smaller than a given value freely assigned. 
These {\em energy-wells} are located via unbiased random
insertion over the whole configurational space and are sampled using
the so called \hr algorithm, which uniformly samples compact regions
of any shape immersed in a space of arbitrary dimensions.  Because 
the bias is defined in terms of the energy landscape it can be
{\em exactly} corrected to obtain the unbiased distribution.
The test-particle energy distribution is then
combined with the Bennett relation for the evaluation of the chemical
potential.  We apply this protocol to a system with relatively small
probability of low-energy test-particle insertion, liquid argon at
high density and low temperature, and show that the energy-biased
Bennett method is around five times more efficient than the standard
Bennett method.
A similar performance gain is observed in the reconstruction of the energy distribution.
\end{abstract}

\maketitle
%PACS, the Physics and Astronomy

\section{Introduction
\label{sec.intro}}

The chemical potential is a central quantity underpinning many
physical and chemical processes, such as phase equilibria, osmosis,
thermodynamic stability, binging affinity and so on
\cite{lu03}. However, its evaluation by computer simulation is more
complicated and time-consuming than for other intensive thermodynamic
quantities, such as the pressure $P$ or temperature $T$.  While $P$
and $T$ can be evaluated from averages over mechanical properties of
molecules (forces, velocities and positions), the chemical potential
is a thermal average and therefore it requires sampling the phase
space of the system.  Indeed, computing the chemical potential is a
special case of the more general problem of computing a free-energy
difference $A_1-A_0$ between two states (labelled as 0 and 1), a
problem for which the inherent difficulty is well understood
\cite{allen,frenkel.book,kollman93,lu03}.  
Free energy perturbation (FEP) is an important category of methods for free energy
calculation; we refer to the recent works by Lu {\it et al.} \cite{lu03} and by Shirts
and Pande \cite{Shirts05} for review and comparisons. As explained by
Lu {\it et al.}  \cite{lu03}, the general working equation for FEP
methods can be cast as
 \begin{equation}
\label{freedif}
  \exp[-\beta(A_1-A_0)]=\frac{ \langle w(u) \exp[-\beta u/2]\rangle_0 }{\langle w(u) \exp[-\beta u/2]\rangle_1},
 \end{equation}
with $\beta =1/k_BT$ and $u\equiv U_1-U_0$ the energy difference between both
systems; $K_B$ is the Boltzmann constant. The angular brackets denote ensemble averages performed on the system
labelled by the subscript ``0'' or ``1''.  The weighting function $w(u)$ is
arbitrary and differs for each method introduced in the literature.

The chemical potential is the free energy difference between two thermodynamic
states differing by the presence of a single molecule.  In other words, the chemical potential is
$A_1 -A_0$ where $A_1=A(N+1,V,T)$ and $A_0=A(N,V,T)$.  Here $A(N,V,T)$ is the
Helmholtz free energy of the system which depends on the number of molecules N,
the volume $V$ and temperature of the system. In order to express the averages
of Eq. (\ref{freedif}) in terms of one-dimensional integrals of the energy
difference $u$ one can then introduce the following distribution functions
\cite{dei89}
\begin{eqnarray}
\label{fg}
f(u)&=& \int \langle \delta \left (u - U_1 +U_0 \right ) \rangle_0 V^{-1} d{\bf r},\\
g(u)&=& \langle \delta \left (u - U_1 + U_0 \right ) \rangle_1,
\label{fg2}
\end{eqnarray}
where $\delta(.)$ is the Dirac delta function. 
In Eq. (\ref{fg}), $U_1=U_1({\bf R}^N, {\bf r})$, where ${\bf R}^N$ is the
configuration of the first N molecules and ${\bf r}$ denotes the configuration
of the N+1 molecule.  Note that in Eq. (\ref{fg}) the N+1 molecule acts as a
``test-molecule'' which probes the system ``0'' (i.e. the system with N
molecules), but does no interact with it. Therefore $f(u)$ is the probability
density of the N molecule ensemble increasing in potential energy by an amount
$u$ if this test-molecule were randomly inserted into the ensemble.  Conversely,
$g(u)$ is the probability density of the (N+1)-molecule ensemble decreasing in
potential energy by an amount $u$ if a randomly selected real molecule were
removed from the ensemble.

>From Eq. (\ref{freedif})-(\ref{fg2}) an expression for the excess chemical
potential $\mu= A_1-A_0 -\mu_{id}$ (where $\mu_{id}$ is the ideal gas chemical
potential \cite{frenkel.book}) can be derived in terms of the $f$ and $g$
distributions \cite{shing82,dei89,lu03}
\begin{equation}
\exp(\beta \mu) =\frac{\int w(u) g(u)  du} {\int w(u) f(u) \exp(-\beta u) du}.
\label{sh2}
\end{equation}
A good choice of the weighting function $w(u)$ is key for the efficiency of
the method. For instance, the Widom method \cite{frenkel.book,allen} ($w(u)=1$) is known to provide 
very poor convergence at large densities. The Widom method is a single stage
FEP, meaning that sampling is only performed in the reference system ``0''
(i.e., in the $f$ distribution, see Eq. (\ref{sh2})).  As discussed by Lu {\it
et al.} \cite{lu03}, multiple staging provides much better efficiency. The
efficiency is generally defined as the reciprocal of the product of the
variance of the estimator multiplied by its cost $n_{cost}$ (that is, the total
number of energy evaluations performed by the algorithm)
\begin{equation}
\varepsilon=(n_{cost} \var[\beta\mu])^{-1}.\label{efficiency}
\end{equation}
Bennett \cite{bennett76} showed that the variance of Eq. (\ref{sh2}) is minimised
if the weighting function is $w(u)=\ferm[\beta(u -c)]$, where
$\ferm(x)=1/(1+\exp(x))$ is the Fermi function and $c$ is an arbitrary
constant. The Bennett estimator is then
\begin{equation}
\beta \mu =\ln\left(\frac{\la \ferm[-\beta(u-c)]\ra_g}{\la\ferm[\beta(u-c)] \ra_f}\right) + \beta c,
\label{bennett}
\end{equation}
where the subscripts $g$ and $f$ indicate (simple) averages over the
distributions $g(u)$ and $f(u)$.
The value of $c$ providing the minimum variance and
maximum overlap is $c=\mu$ and to evaluate $\mu$ using the optimum $c(=\mu)$
one requires to use a self-consistent procedure, iterating the value of
$c$ in Eq. (\ref{bennett}) and resetting $c=\mu$ until $\la
\ferm[-\beta(u-c)]\ra_g=\la \ferm[\beta(u-c)]\ra_f$.  In practise, this step
only requires a small number of iterations.  Recent publications \cite{lu03,Shirts05} 
demonstrate that the Bennett method remains the best general method to compute the chemical potential for many applications.

Note that the Bennett method is a two-stage FEP and therefore it also requires 
sampling of the system ``1''.  In the case of the determination of the chemical potential this
system has N+1 molecules and $g(u)$ is obtained from its single-molecule
energy distribution.  However this extra requirement is not really a drawback.
Lu {\em et al.} \cite{lu03} showed that, provided $N >O(100)$, the
$g-$average can be evaluated in the same simulation as is used to sample the $f$
distribution (system ``0'') without any noticeable loss in accuracy.  The $g$
distribution (constructed from the energy of the real particles) is thus a
byproduct of the simulation so the average $\la\ferm\ra_g$ does
not demand any extra computational cost. 

Another group of methods for determination of the chemical potential are based on {\em biased}
instead of uniform sampling.  In particular, cavity-biased methods
first select spherical cavities of minimum radius $R_c$ (a free parameter) in which
to insert the test-molecule.  This accelerates the evaluation of the ensemble
average in dense phases because the low-energy configurations of the
test-molecule (with large Boltzmann factors) are usually located in larger
cavities with less steric hindrance.  Variations of this method have been
proposed by several authors; these include the Cavity Insertion Widom method (CIW) due to Mezei
and coworkers \cite{jed00}, the Excluded Volume Map Sampling by Deitrick {\it
et al.} \cite{dei89} and the method proposed by Pohorille and Wilson
\cite{poh96}. The cavities are located by a grid search over the whole
simulation cell. A cavity centre is assigned at each grid point whose distance
to the closest particle is greater than $R_c$. In order to correct the bias
introduced in sampling only inside the cavities one also has to calculate the
probability of finding a cavity, which is obtained in the same grid-search
step.  A drawback of the cavity-biased method is that it is only indirectly  related to the
test-particle energy via the excluded
volume.  This fact introduces a certain inaccuracy in the estimation of the
chemical potential, as it can depend on the value of the cavity radius $R_c$
selected. For instance, the CIW has recently been used to calculate the
chemical potential of several species across a lipid bilayer \cite{jed00}.  As
a test calculation the authors estimated the chemical potential of
water in water and reported variations of about 1 Kcal/mol as $R_c$ was varied
from $2.6\AA$ to $2.8\AA$. Also, using $R_c\in [2.6, 2.9] \AA$ resulted in
uncertainties of about 2 Kcal/mol in estimates of the excess chemical
potential of some species across the lipid layer.  Note that the important
region of the cavity-biased method is constructed over the translational degrees
of freedom of a ``coarse-grained'' spherical molecule with an effective
radius. This means that it can only be applied to small solutes with spherical
or roughly spherical shapes \cite{dei89}.

In this work we present an energy-biased method for the estimation of the
chemical potential and reconstruction of the energy distribution $f(u)$ in
dense phases. The idea is to restrict the sample to an important region
defined by the set of bounded domains in the configurational space of the
test-molecule where the energy $u$ is smaller than a given free parameter
$u_{w}$.  We denote as an {\em energy-well} each compact subdomain within the test-molecule
energy-landscape for which $u<u_{w}$. Note that the present approach retains the
main benefit of the cavity-biased method, but provides an exact
evaluation of the energy distribution $f(u)$ and the chemical potential,
because the energy-wells are defined directly in terms of the energy landscape. Moreover our energy-biased method
does not assume any particular molecular shape and therefore it may be used for
non-spherical molecules and can coherently sample over rotational degrees of
freedom as well. 

We also note that the number of stages are not limited to two. When 
systems 0 and 1 are very different it may be impossible within the 
simulation time to sample the importance region of the two systems. In 
this case it is more efficient to compute the total free energy 
difference by using a set of intermediate states. The energy bias method 
can be applied  on each of these intermediate state transitions at the 
cost of performing independent simulations for each state. Other  
approaches include, for instance, slow and fast growth methods where the 
system is changed from one state to another within a certain simulation 
time $\tau$ (large for slow growth). The fast growth method consists of 
sampling rapid transformation from many simulations which are then 
combined by using Jarzynski  nonequilibrium work relation \cite{jarz97} to 
obtain the total free energy difference.

The rest of the paper proceeds as follows. The energy-biased method is explained
in Sec. \ref{sec.method}, while in Sec. \ref{sec.optimal} we derive an analytical
expression for the efficiency of the method and estimate the optimal parameter
$u_{w}$ by maximising the efficiency. In Sec \ref{sec.results} the method is tested in liquid argon at
high density (modelled as Lennard-Jones atoms) where it is used to reconstruct the
test-particle energy distribution $f(u)$ and the chemical potential. We
also demonstrate the gain in efficiency obtained with energy-biased sampling with
respect to uniform sampling. We conclude with a summary of our findings in Sec. \ref{sec.con}. Finally in 
Appendix \ref{hitandrunapp} we briefly explain the \hr algorithm which efficiently samples
bounded regions of arbitrary shape immersed in an arbitrary number of dimensions.

\section{Overview of the method \label{sec.method}} 

As stated in the introduction,  energy-biased sampling consists of
uniform sampling of the importance region defined by the set of
subdomains in the test-molecule configurational space where its
potential energy is less than $u_w$.  The probability density is
therefore given by
\begin{equation}
\label{fb}
h(u)=\left\{
\begin{array}{cc}
  f(u)/\fstar & u\leq u_{w} \\
0 & u> u_{w}, \\
\end{array}
\right.
\end{equation}
where the normalisation factor $\fstar\equiv\int_{-\infty}^{u_{w}} f(u) du$ is
the cumulative probability of the unbiased distribution $f(u)$ and $u_{w}$ is
an arbitrary energy (free parameter). 

Note that the energy-biased distribution of Eq. (\ref{fb}) 
can be straightforwardly combined with any of the popular
methods to calculate the chemical potential from Eq. (\ref{sh2}).  We shall
use the Bennett method due to its excellent performance. Introducing the
weighting function $w(u)=\ferm[\beta(c-u)]$ in Eq. (\ref{bennett}) and using
Eq. (\ref{fb}), one obtains the energy-biased Bennett estimator for $\beta
\mu$,
\begin{equation}
\label{general}
\beta \mu = \ln\left(\frac{\la \ferm_c \ra_g}{\fstar \fermac} \right) + \beta c,
\end{equation}
where we have introduced the notation $\ferm_c \equiv \ferm[\beta(u-c)] $ to
indicate that after the ensemble average we still have a function of $c$.  As
before, the subscript $h$ indicates the average over the biased
distribution of Eq. (\ref{fb}).

Sampling from the energy probability distribution $h(u)$ requires a 
more  careful consideration of the energy landscape of the system.
We indicate by ${\bf r}$ a configuration of the (N+1)th
molecule and by ${\bf R}$ the configuration 
of the remaining N molecules.  For a simple argon fluid
${\bf r} \in D$ where $D \subset R^3$, while for a 3 sites flexible water model like TIP3P
$D \subset R^9$,  which includes the three Euler angles determining the molecule orientation, the H-O-H angle and the
two H-O distances.

%%%% INCLUDE FIG1 HERE

\begin{figure}[tb]
\begin{center}
\centerline{\psfig{figure=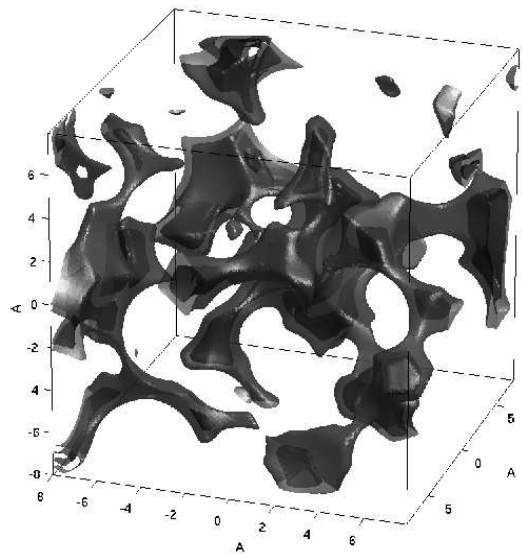,width=7cm}}
\end{center}
\caption{Energy landscape for the three-dimensional configurational space generated by 
 inserting an argon atom in a cube of side $16$ {\AA} of argon fluid.  The
isosurfaces of regions $A_{u_{w}}$ are shown for $u_{w}$ equal to 1 (dark grey) and to $10$ Kcal/mol (light grey).
}
\label{figg1}
\end{figure}

As shown in Fig. (\ref{figg1}), the region 
\begin{equation}
 A_{u_{w}} =  \{ {\bf r}\in D : u({\bf r,R})<u_{w}  \}
\end{equation} 
 is composed of many disconnected bounded regions of different sizes such that
$A_{u_{w}}=\cup_{\alpha} A_{u_{w}}^{\alpha}$, where each $A_{u_{w}}^{\alpha}$
is now a connected region.  Of course, for $u_{w} \rightarrow \infty$ we have
that all the regions $A_{u_{w}}^{\alpha} $ connect and $A_{\infty}^{\alpha}=D
$, the entire domain.  The sampling algorithm must reproduce a uniform probability
distribution
\begin{equation}
p_{u_{w}}({\bf r}) = \frac{1}{\vol(A_{u_{w}})},\label{lambdaprob}
\end{equation}
where $\vol(A_{u_{w}})$  is the volume of the region.

For a given energy bias $u_{w}$, the algorithm for selecting configurations ${\bf
r}$ according to Eq. (\ref{lambdaprob}) can be described in terms of two main
steps which are applied iteratively:
\begin{enumerate}
\item Locate  a compact energy-well $A_{u_{w}}^{\alpha} $ in the configurational
space D, where $u<u_{w}$.
\item Sample the energy-well $A_{u_{w}}^{\alpha}$ with a uniform probability 
density.
\end{enumerate}

The simplest procedure for locating energy wells in step (1) is to perform a
random search over the whole configurational space until a fixed number of
cavities is found. This procedure, however, does not avoid the probability of
exploring the same well more than once, and we observed that it can easily
lead to highly correlated data.  Instead we perform step (1) by choosing points
on a grid within the whole configurational space of the test-molecule.  In the
case of the Lennard-Jones fluid, the three-dimensional configurational space
is probed at the nodes of a Cartesian grid of size $n_x\times n_y\times n_z$,
where $n_{\alpha}$ is the number of nodes along the coordinate $\alpha$.  We
observed that the minimum distance between nodes that guarantees statistically
independent samples is around $0.5\sigma$.

An energy well is found
at each node where the energy of the test-molecule is
$u<u_{w}$. Then, the locations of each of these nodes are used as starting
configurations for independent well samplings.
In this way we ensure that we
are sampling different cavities 
for each explored configuration (snapshot) of
the system.  Note that using grid-sampling the number of cavities found per
snapshot is a fluctuating quantity.

The search requires an average of $n_{0}=1/\fstar$ energy evaluations to
locate one well (i.e. one configuration with energy $u<u_{w}$.)  
During this same step (1) one can calculate the cumulative probability
$\fstar$ from the estimator $m/n_0$, with $n_0$ being the total number
of samples (Bernoulli trials) and $m$ the number of successful trials 
with $u<u_{w}$, i.e., the total number of energy-wells found. 
This number $m/n_0$ converges to $\fstar$ as $n_0 \rightarrow \infty$
and, for a finite number of statistically independent trials $n_0$, its
variance is $(1-\fstar)\fstar/n_0$.
 In practise, the estimation of $F_w$ 
requires the number of unbiased samples to be $n_0>>1/F_w$; this condition
also ensures that a significant number of energy-wells ($m>0$) are to be found.

Step (2) of the loop mentioned above requires a procedure to sample in
an unbiased way the interior of each energy well.  This is a delicate step
because any bias incurred in sampling the importance region will be transfered
to the estimator for $\beta \mu$, resulting in inaccuracy of the method.
To tackle this problem we use the so-called \hr algorithm \cite{Smith84},
which is explained in Appendix \ref{hitandrunapp}.

\section{Efficiency and optimal parameters of the method \label{sec.optimal}}

We now calculate the efficiency of the method and provide a way of choosing
the optimal value of the parameter $u_{w}$ by maximising the efficiency.  We
also compare the efficiency of the estimator in Eq. (\ref{general}) based on
energy-biased sampling with that of the standard Bennett algorithm
of Eq. (\ref{bennett}).

\subsection{Energy-biased Bennett method}

The variance of the Bennett method can be cast in terms of the probability
densities $f(u)$ and $g(u)$. Starting from Eq. (\ref{bennett}), after some algebra the variance
of the Bennett method assumes the form
\begin{equation}
  \label{vb2}
\var_{B}[\beta \mu] = \frac{1}{n_0 \la \ferm[\beta(u-c)] \ra_f },
\end{equation}
where $n_0$ is the number of insertions used to sample the complete
configurational space of the test-particle. Note that the computational cost of
the standard Bennett method is $n_0$, so according to
(\ref{efficiency}) and Eq. (\ref{vb2}) its maximum efficiency is given by
\begin{equation}
\label{efbenet}
  \varepsilon_{B} = \la \ferm_c \ra_f.
\end{equation}

Let us now consider the variance of the estimator in Eq. (\ref{general}), which is
the sum of the variance of the estimator for $\fstar$ and the estimator for
the ensemble average
\begin{equation}
  \label{variance}
  \var_{EB}[\beta \mu] = \var[\ln \fstar] + \frac{1}{n_{w} \fermac} \simeq 
  \frac{1}{n_0 \fstar} + \frac{1}{n_{w} \fermac },
\end{equation}
where we have used the relation 
$\var[\ln(\fstar)] \simeq \var[\fstar]/\fstar^2= (1-\fstar)/(n_0 \fstar)\simeq 1/(n_0 \fstar)$,
for $\fstar<<1$.  Here $n_0$ is the number of random insertions in the entire 
configurational space and $n_{w}$ is the number of independent samples within
the importance region $u<u_{w}$.

The probability of finding an energy-well with $u<u_{w}$ using uniform
sampling over the whole configurational space is $\fstar$, so the
number of cavities found after $n_0$ trials is $m=\fstar n_0$.  If the
number of statistically independent samples per well is $s$, the total
number of independent samples within the restricted configurational
space $u<u_{w}$ is
\begin{equation}
\label{n1}
  n_{w}=n_0 s\fstar. 
\end{equation}
We note that the number of independent samples per well $s$ depends on
the fluid considered and, of course, on the biasing energy $u_{w}$.  In
Appendix \ref{numindsample} we provide a way of estimating $s$ from the
outcome of the data obtained from \hr sampling. Inserting
Eq. (\ref{n1}) into Eq. (\ref{variance}) one obtains for the energy-biased
algorithm
\begin{equation}
\label{var2}
 \var_{EB}[\beta \mu] = \frac{1}{n_0} \left( \frac{1}{\fstar} + \frac{1}{s\ferma } \right).
\end{equation}
In deriving Eq.(\ref{var2}) we used that $\ferma = \fstar\fermac$ up to a
negligible amount. This can be seen by noticing that the function $\ferm[\beta(u-c)]$ in the integrand 
of $\ferma=\int_{-\infty}^{\infty} f(u) \ferm[\beta(u-c)] du$ decays exponentially for $u>c$.
Hence, in any practical case ($u_w>c$) most of the integral weight comes from $u<u_w$,
for which the energy-biased reconstruction of the energy profile $f(u)$ is
exact (see Fig. \ref{figg3}).

We now evaluate the cost, which is given by the total number of energy
evaluations of the test molecule needed to obtain $n_w$ samples:
\begin{equation}
  \label{cost}
n_{cost}=  n_0 + n_{w}/\ac,
\end{equation}
where $\ac<1$ is the acceptance ratio of the \hr sampling
algorithm, defined in Appendix \ref{hitandrunapp}. 
Introducing Eq.(\ref{n1}) into Eq. (\ref{cost}) we obtain
\begin{equation}
\label{cost2}
  n_{cost}= n_0\left(1+\frac{s\fstar}{\ac}\right).
\end{equation}

 For the energy-biased algorithm the efficiency is
 $\varepsilon=(n_{cost}\var_{EB}[\beta\mu])^{-1}$.  Using Eq.(\ref{var2}) and
 Eq.(\ref{cost2}) one obtains
\begin{equation}
\label{eff}
 \varepsilon_{EB}^{-1}=\frac{1}{\fstar}+\frac{1}{s \ferma}+ \frac{s}{\ac}+ \frac{\fstar}{\ac\ferma}.
\end{equation}
By maximising the efficiency $\varepsilon=\varepsilon(\fstar)$ in Eq. (\ref{eff})
with respect to $\fstar$, one obtains the optimal value $\fstar^{opt}$ and the
maximum efficiency $\varepsilon_{EB_{\max}}=\varepsilon_{EB}(\fstar^{opt})$:
\begin{eqnarray}
  \label{fopt}
\fstar^{opt}&=&\sqrt{\ac \ferma}\\
\label{emax}
\varepsilon_{EB_{\max}}^{-1}&=& 2 \frac{1}{\sqrt{\ac \ferma}} + \frac{s}{a} +\frac{1}{s\ferma}.
\end{eqnarray}
Finally, we compare the efficiency of the energy-biased algorithm with that
provided by the Bennett algorithm, given by $\varepsilon_{B}=\ferma$.  According
to Eq. (\ref{emax}) the ratio of efficiencies is given by
\begin{equation}
  \label{effratio}
\frac{\varepsilon_{B}}{\varepsilon_{EB_{\max}}}= 2 \sqrt{\frac{\ferma}{\ac}} +\frac{s\ferma}{\ac}+ \frac{1}{s}.
\end{equation}
Equation (\ref{effratio}) yields the range of values of $\ferma$ for which the 
energy-biased Bennett estimator for $\beta \mu$  method is more 
efficient than the standard (unbiased) Bennett algorithm. Note that for $s=\sqrt{\ac/\ferma}$
the efficiency ratio given by
Eq. (\ref{effratio}) reaches its minimum value, $\varepsilon_{B}/\varepsilon_{EB_{\max}}= 4\sqrt{\ferma/\ac}$,
and therefore $\varepsilon_{B}<\varepsilon_{EB}$ if $\ferma>\ac/16$. 
Hence the energy-biased method is suited for fluids at high densities or low temperatures or for 
molecular fluids with low insertion probability. In this regime $\ferma<<\ac/16$
and the dominant term in Eq. (\ref{effratio}) is $1/s$, hence
$\varepsilon_{EB_{\max}}\simeq s \varepsilon_{B}$. In other words, the maximal efficiency
of the present energy-biased method is limited by the average number $s$ of
independent samples that can be obtained within one energy-well. As shown in
 Appendix \ref{numindsample}, for the Lennard-Jones fluid we have observed
that in the most unfavourable case (high density and low temperature) $s\sim [5-10]$.

\subsection{Reconstruction of the energy distribution}

We now show that the reconstruction of $f(u)$
using the energy-biased procedure (EB) is faster and more efficient than 
that obtained using any unbiased sampler which uniformly explores the whole configurational
space. To that end we consider the evaluation of the cumulative probability
$F(u)=\int_{-\infty}^u f(u')du$ for $u<u_{w}$ (i.e. for $F(u)<\fstar$).  We
shall compare the variance of two estimators for $F$: one based on uniform
insertion over the whole domain and the other based on the energy-biased
procedure. The variance of the unbiased estimator is simply
$\var(F)=F(1-F)/n_0$ and for low energies ($F<<1$) its efficiency is
$1/F$. The expected value of the energy-biased estimator is $H \fstar$,
where $H(u)=\int_{-\infty}^u h(u')du'$ is the cumulative
probability of the biased distribution in Eq. (\ref{fb}).  This estimator is
constructed as a product of two statistically independent fluctuating
variables and its variance is \cite{goodman60}
\begin{eqnarray}
  \label{varF}
\var_{EB}(F)&=&\var(H\fstar)=\fstar^2 \var(H) \\
\nonumber &+& H^2 \var(\fstar) + \var(\fstar)\var(H).
\end{eqnarray}
Using $\var(H)=H(1-H)/n_{w}$ and $\var(\fstar)=\fstar(\fstar-1)/n_0$   one  obtains
\begin{equation}
  \var_{EB}=\frac{\fstar(\fstar-1) H^2}{n_0}+\frac{H(1-H) \fstar^2}{n_{w}}+\frac{\fstar H(1-H)}{n_0n_{w}}.
\end{equation}
Note that, as expected, for $H\simeq 1$ one recovers the variance of the
unbiased insertion method.  The interesting part of the energy distribution is
the importance region, located in the low energy range, where $H<<1$.
In this regime one can make the approximation $1-H\simeq 1$. Using
$F=H\fstar$ and $n_{w}=n_0s\fstar$, one gets
\begin{equation}
\label{var_reconst}
  \var_{EB}=\frac{F}{n_0}\left(\frac{F}{\fstar}+\frac{1}{s}+\frac{1}{n_0 s\fstar}\right).
\end{equation}
Note that the term in brackets is the reduction in variance with respect to
uniform unbiased sampling.  Because $\fstar$ is evaluated from $n_0$ probes,
this means that necessarily $n_0>> 1/\fstar$ so the third term inside the
brackets is much smaller than unity. On the other hand, for the low energy
range considered $F<<\fstar$ and one finally concludes that $\var_{EB}\simeq
\var(F)/s$, where $\var(F)\simeq F/n_0$ is the variance obtained in the
unbiased uniform sampling of the whole domain.

The cost associated with the energy-biased procedure is
$n_{cost}=n_0(1+s\fstar/\ac)$.  In the case of a Lennard-Jones liquid we have found
that $\ac\simeq 0.17$ and $s\sim O(10)$, while the optimal cumulative
probability is $\fstar \lesssim 10^{-3}$. This means that, in practical
situations, $s\fstar/\ac \lesssim 1$ and $n_{cost}\gtrsim n_0$.  Thus,
according to Eq. (\ref{var_reconst}) the energy-biased sampling procedure is
around $s$ times faster than a uniform unbiased (grid or random) sampler
in reconstructing the low energy range of $f(u)$. As before,
$s$ is the average number of independent samples taken per well.

\section{Results \label{sec.results}}

In order to confirm the foregoing theoretical relations about efficiency and variance
reduction, we performed molecular dynamics simulations of a Lennard-Jones
liquid at high density and low temperature ($\rho=0.0236\AA^{-3}$ and
$T=84$K). These simulations were performed in a cubic periodic box of side
$L=10\sigma$.  We used the standard Verlet method \cite{allen} to integrate
Newton's equations of motion, incorporating a Langevin thermostat \cite{grest1} to keep
the system in the NVT ensemble.

During the simulation, the iterative loop (1)+(2) explained in Sec. \ref{sec.method} was
performed  $m$ times per time interval $\delta t_{samp}=0.5\tau$,
which corresponds to about three times the collision time.  The search for wells performed
in step (1) was done  by probing at the nodes of a Cartesian grid comprising $15^3$
nodes.  This ensured that the explored cavities are independent. 
All the cavities found in step (1) were sampled using the \hr
algorithm (see Appendix \ref{hitandrunapp}).

\subsection{Estimation of the chemical potential}

\begin{figure}[h]
\includegraphics[width=8cm,totalheight=12cm]{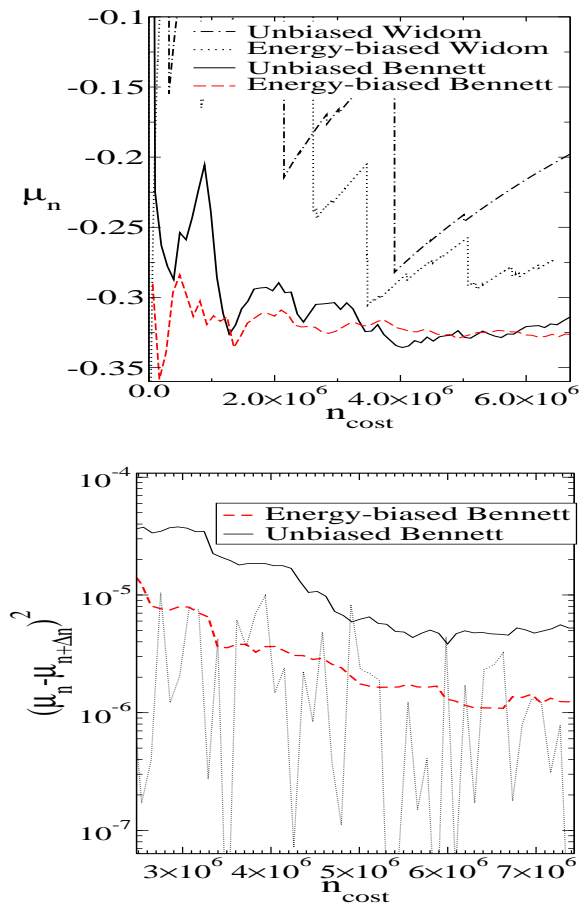}
\caption{(Top) The estimation of the chemical potential plotted
against the overall number of energy probes ($n_{cost}$). We compare the standard
(uniform sampling) Bennet and Widom methods with the corresponding energy-biased versions of these methods.
The calculations correspond to a Lennard-Jones fluid with $\rho=0.0236\AA^{-3}$, and $T=84$K
($\rho=0.92\sigma^3$ and $T=0.7 \epsilon/k_B$ in reduced LJ units);
the energy-biased sampling was done using $u_{w}=14.19$ Kcal/mol and $d=15$ samples per well.
(Bottom) The convergence measured as the squared difference between consecutive 
estimations  with increasing cost ($\Delta n=10^5$).
In the energy-biased method the cost is given by $n_{cost}=n_0\left(1+d\fstar/\ac\right)$,
where $n_0$ is the number of random probes used to evaluate 
$\fstar(=F(u_{w})=0.00122)$ and $\ac=0.165$ is the acceptance ratio of the \hr sampler. 
Circles are the successive values of $(\mu_n - \mu_{n+\Delta n})^2$ (shown only for the energy-biased case)
and solid lines are the average over twenty consecutive differences.
}
\label{figg2}
\end{figure}

One way to measure the efficiency of the method is to evaluate the
convergence of the estimated value of the chemical potential for an
increasing number of test-particle probes $n_{cost}$.  Convergence can
be calculated from the difference between successive values of
$\mu_n$, where $n(=n_{cost})$ indicates the total number of
evaluations of the test-particle energy. Figure \ref{figg2} shows how
this difference decreases in calculations based on both the
energy-biased and the unbiased samples. These calculations correspond
to liquid  argon with number density $\rho=0.0236 \AA^{-3}$ and
temperature $T=84$K (these values correspond to $\rho=0.92\sigma^3$
and $T=0.7$ in Lennard-Jones units), for which the average of the
Fermi function is $\ferma=8.9\times 10^{-6}$.  According to
Eq. (\ref{fopt}) the optimum value of $\fstar$ is $0.0012$, which
corresponds to $u_{w}\simeq 14.19$ Kcal/mol. We selected the predicted
optimum parameter ($u_{w}=14.19$ Kcal/mol) and performed $d=15$
samples per well. As can be seen in Fig. \ref{figg2}, for equal numbers
of energy probes ($n=n_{cost}$), the average
difference between successive estimates of the chemical potential via
the energy-biased method is about five times smaller than that obtained
with the unbiased sampler.  As predicted by Eq. (\ref{effratio}), such
a gain in efficiency is consistent with the average number $s$ of
independent samples per well (see table \ref{table_s}), which for this
simulation was $s\simeq 5$.

Evaluations of the chemical potential for Lennard-Jones (LJ) fluids 
are shown in Table \ref{table_mu} together with the estimated efficiency of
each calculation. For a LJ fluid with  $\rho=0.02360 \AA^{-3}$ and $T=84$K 
the numerically obtained net gain is around 7, which coincides with the prediction 
in Eq. (\ref{effratio}) using $s=7$. For illustrative purposes
we also analysed a case for which the efficiency of our implementation of the energy-biased sampling
is similar to the uniform-unbiased Bennett method. For instance,  $\ferma=0.0102$ for
$\rho = 0.01755 \AA^{-3}$ and $T=178.5$K.
Using $\ac=0.165$ and the (optimum) number of samples $s=\sqrt{\ac/\ferma}\simeq 4$ in Eq. (\ref{effratio})
one obtains $\varepsilon_{B}/\varepsilon_{EB_{\max}} \simeq 1$; our numerical
calculations, with $u_w=7.33$ and $d=8$, confirmed this conclusion. 
We note that for any value of $u_{w}$ considered the
energy-biased estimation of the chemical potential $\mu$  agrees within about
$0.01$ Kcal/mol with the unbiased Bennett result. This is
illustrated in Table \ref{table_s} where we show the estimated $\mu$
for the higher density liquid, using several values of $u_{w}$.

\begin{table*}
  \begin{tabular}{|c|c|c|c|c|c|c||c|c|c||r|r|}
\hline
$\rho$ ($\AA^{-3}$) & $T$ (K) & $\mu_{EB}$& $\mu_{B}$& $\varepsilon_B^{-1}$& $\varepsilon_{EB}^{-1}$ & $\varepsilon_{EB}/\varepsilon_{B}$ & $F_w$ & $d$ \\
\hline
0.02360 & 84    & -0.336 & -0.323  & $0.9\times 10^{5}$  & $(1.2 \pm 0.1)\times 10^{4}$ & $7 \pm 1$ & 0.00122 & 15     \\
\hline
  \end{tabular}
\caption{Comparison of the chemical potential (in Kcal/mol) calculated via the standard Bennett method 
(i.e. using uniform unbiased sampling) and the energy-biased Bennett (subscript $EB$). The inefficiency of both methods 
(reciprocal of efficiency) is also shown. In the case of the standard Bennett method we write the minimum inefficiency
($\varepsilon_B^{-1}=\la \ferm_{\mu}\ra_f$) while the inefficiency of the energy-biased method was obtained from 
numerical calculation of the variance of $\beta \mu$, using block-analysis 
(see Appendix B or e.g. Refs. \cite{allen,frenkel.book}) and agrees within error bars with 
the theoretical expression of Eq. (\ref{eff}) (see text).
The error in $\varepsilon_{EB}/\varepsilon_{B}$ comes mainly from the uncertainty in the 
numerical calculation of $\var_{EB}$.}
\label{table_mu}
\end{table*}

%%%% INCLUDE FIG2 HERE

\subsection{Reconstruction of the energy distribution $f(u)$}

%%%% INCLUDE FIG3 HERE
\begin{figure}[h]
\includegraphics[width=8cm,totalheight=8cm]{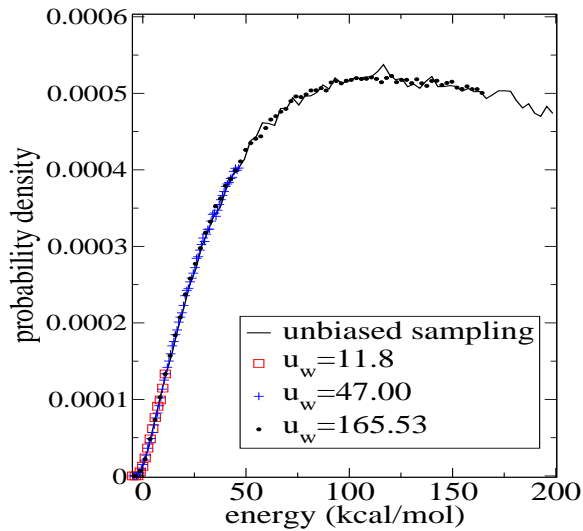}
\caption{The energy distribution $f(u)$ 
obtained from $4\times 10^6$ random insertions over the whole configurational domain is compared
with  energy-biased sampling in the restricted configurational
space $u<u_{w}$.  The calculations correspond to the same case as in Fig. \ref{figg2}.
The energy cavities are sampled using the \hr algorithm, which provides an unbiased reconstruction of the energy
distribution for any value of $u_{w}$ chosen.}
\label{figg3}
\end{figure}

%%%% INCLUDE FIG4 HERE
\begin{figure}[h]
\includegraphics[width=8cm,totalheight=8cm]{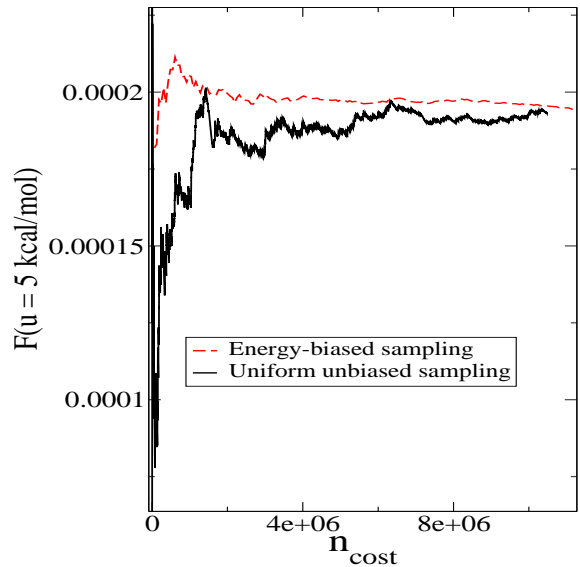}
\caption{The cumulative probability $F(u)=\int_{-\infty}^u f(u') du'$ for
$u=5$ Kcal/mol {\em versus} the total number of energy evaluations of the
test-particle $n_{cost}$. The liquid is the same as in
Fig. \ref{figg2}. We compare the estimations of $F(u)$
for grid-sampling (with a regular mesh of $36^3$ nodes) and for energy-biased
samplings within $u<u_{w}=14.19$ Kcal/mol, performing $d=15$ samples per well.
The cumulative probability at $u_w$ is $\fstar=F(u_{w})=0.00122$.
}
\label{figg4}
\end{figure}

In Fig. \ref{figg3} we compare the reconstructed energy distribution $f(u)$
at energies $u<u_{w}$ with that computed from an unbiased method, which
consists of a large number of random insertions within the entire configurational
space.  Figure \ref{figg3} clearly illustrates that the energy-biased method
{\em exactly} reproduces the unbiased distribution $f(u)$ for energies smaller
that $u_{w}$.  This attractive feature is a consequence of the fact that it is easy to
exactly correct for the bias in terms of the cavity energies. This is not true 
for the accessible volume of the molecule, as  in cavity-biased
procedures \cite{jed00,dei89}.

In order to illustrate the above conclusion we show in Fig. \ref{figg4}
the estimation of the cumulative probability $F(u)$ {\em versus} the total number of
test-particle energy probes used for the evaluation. The particular case shown 
corresponds to $u=5$ Kcal/mol, for a LJ liquid at $\rho=0.0236\AA^{-3}$ and
$T=84$K.  The energy-biased sampling was done using $u_{w}=14.19$ Kcal/mol and
$d=15$ samples per well, and for this calculation we obtained $s\simeq 5$ 
(see Appendix and Table \ref{table_s}).  Compared with the unbiased
procedure, the reduction of variance provided by the energy-biased sampler is
immediately apparent on inspection of Fig. \ref{figg4}.  A numerical evaluation of
the variance of each data set in Fig. \ref{figg4} provides: $\var_{EB}=
4.14\times 10^{-5}/n_0$, while the (best) result for the algorithm based on uniform
unbiased sampling is $\var(F)=F/n_0=1.9\times 10^{-4}/n_0$. Hence the net gain in
efficiency is about 4.6, in agreement with the value of $s=5$ obtained from
the independent correlation analysis explained in Appendix \ref{numindsample}.
As shown in Table \ref{table_mu}, the estimated net gain in the evaluation of the chemical potential
compared with the unbiased Bennett method is $7\pm 1$, which is close to the estimate 
$s\simeq 5 $ obtained from the analysis of the cumulative probability.

\section{Conclusion \label{sec.con}}

We have presented a new method for sampling the energy of a
test-molecule in order to calculate single-particle ensemble averages
and, in particular, the chemical potential. The method, called {\it
energy-biased} sampling, restricts the important region to the bounded
domains in the test-molecule energy-landscape where the test-molecule
energy $u$ is smaller than a given free parameter $u_{w}$.  This
energy-biased sampling retains the principal benefit of cavity-biased
methods \cite{jed00,dei89} in the sense that, by sampling only within
regions with a significant Boltzmann factor, convergence is greatly
accelerated with respect to uniform sampling. Furthermore, because the
energy-biased sampling is accurately defined in terms of the
test-particle energy it has some important benefits: first, it allows
accurate reproduction of the test-particle energy distribution $f(u)$
and the chemical potential; second, it is possible to sample cavities
of arbitrary shape (not only spherical ones) and to generalise the
cavity dimensionality to include the rotational degrees of freedom in
the energy-well reconstruction; finally, and rather importantly, it
enables one to combine the sampling results with standard free energy
perturbation (FEP) formulae.  In particular, we combined it with the
Bennett method \cite{bennett76} which minimises the variance of the
estimator and has proved to be the best method in the
literature\cite{lu03,Shirts05}.  Energy-biased sampling is a general
protocol to bias the sampling and consists of two sequential steps:
(1) searching and (2) sampling the interior of energy-wells. In this
work we have implemented these two steps using relatively simple
algorithms: uniform unbiased search and \hr sampling.  However we note
that other solutions are also possible. 
For instance, non-uniform
sampling of the importance region may surely increase the efficiency
of the present method. In dense systems, the searching step becomes the most difficult one 
and a more effective extension of this method could be to perform a biased search (using, for instance, some variation of the {\sc
usher} algorithm \cite{usher,water_usher}) so as to significantly increase the
probability of finding favourable cavities for insertion of the test
particle. These extensions are
left for future studies.

\acknowledgements This research was supported by the EPSRC Integrative
Biology project GR/S72023 and by the EPSRC RealityGrid project
GR/67699. R.D-B acknowledges support 
from the European Commission via the MERG-CT-2004-006316 grant and 
from the Spanish research grants FIS2004-01934 and CTQ2004-05706/BQU.

\appendix
\section{Sampling bounded regions with the \hr algorithm 
\label{hitandrunapp}}

There exists a relatively 
large literature on sampling a bounded connected region (see for instance Ref. \cite{liu01} and references therein). 
In this work we have used the so-called  \hr algorithm for its simplicity and good performance \cite{liu01}.
The \hr sampler is a special Monte Carlo Markov chain which draws
numbers from an assigned distribution \cite{Smith84,liu01} $p({\bf r})$,
where ${\bf r}\in A$ lies within a bounded connected region of an n-dimensional space $A\subset R^n$.
In our case, $p({\bf r})$  is a uniform probability density over the region $A_{u_{w}}^{\alpha}$ such that
\begin{equation}
p({\bf r})=\frac{1}{\vol(A_{u_{w}}^{\alpha})}.
\end{equation}

The \hr algorithm starts from a point ${\bf r_0}$  within the bounded region $A$ and 
performs the following steps: 
\begin{enumerate}[i.]
  \item Choose a random direction ${\bf e}$ and find the intersections of the cavity border with the line
${\bf r}(\lambda)= {\bf r}_0 + \lambda {\bf e}$, where  $\lambda$ is a real number. As the cavity $A$ is bounded
the intersection is composed by two points ${\bf r}(\lambda^+)$ and ${\bf r}(\lambda^-)$ 
(here $\lambda^+>0$ and $\lambda^- <0$). 
\item Select a point ${\bf r}_1$ within the segment (${\bf r}(\lambda^+)$, ${\bf r}(\lambda^-)$), i.e.,
  \begin{equation}
{\bf r}_{1} = {\bf r}(\lambda^-) + \xi ({\bf r}(\lambda^+)- {\bf r}(\lambda^-))
  \end{equation}
where $\xi\in (0,1)$ is a uniformly distributed random number. 
\item Sample at ${\bf r}_{1}$, set ${\bf r}_1\rightarrow {\bf r}_{0}$ as the new starting point and go to (i).
\end{enumerate}

The above procedure is repeated to obtain the desired number of samples $d$.
In our case the starting point for the sample chain,  ${\bf r}_0$,  is the test-particle configuration returned
by the algorithm for energy-well searching ($U({\bf r}_0,{\bf R})<u_{w}$).
In order to locate the borders of the energy well ${\bf r}(\lambda^+)$ and ${\bf r}(\lambda^-)$ 
we use the following procedure. Starting from ${\bf r}_0$
we cross the well along the line defined by the random unit vector ${\bf e}$  
moving in steps of size $\delta s$, i.e., according to 
\begin{equation}
{\bf r}(k) =  {\bf r}_0 +  \, k\, \delta s \, {\bf e} , \label{walk} 
\end{equation}
with $k$ being an integer starting from $k= \pm 1$. The energy is computed at each point ${\bf r}(k)$ 
until one crosses the edges of the well at $k=k^+$ and $k=k^-$ (for which $u({\bf r({k^{\pm})}},{\bf R}) > u_{w}$).
An approximate location of the cavity borders is provided by setting $\lambda^{\pm}=k^{\pm}$.
We used typically $\delta s   \simeq 0.3\AA$ and required, on average, about five iterations to cross the well 
in one random direction (this value depends on the density and $u_w$). 
Note that the acceptance ratio is $\ac = \la k^+ - k^- \ra^{-1}$ and for the high density cases
considered here $\ac \simeq 0.17$.

\section{Optimal number of sampling directions \label{numindsample}}
It is possible to reduce the cost without increasing the variance by setting
the number of samples per cavity $d$ equal to or somewhat larger than $s$, the
average number of independent samples per cavity.  Note that the number of
statistically independent samples within one cavity is $s=d/\tau_c$, where
$\tau_c$ is an empirically estimated autocorrelation length of the whole chain
of data. This number $\tau_c$ can be estimated from the large $m$ limit of the
quantity $m\,\var[\ferm^{(m)}]/\var[\ferm]$, where $\ferm_c=\ferm[\beta(u-c)]$
is the Fermi function evaluated at a single energy $u$ and $\ferm^{(m)}$
denotes the mean of $m$ consecutive $\ferm$ values.

The value of $s$ can be estimated by performing several \hr samplings with an
increasing number of directions per cavity $d>s$, then computing $\tau_c$
for the chain of samples and evaluating $d/\tau_c$, which should be nearly
independent of $d$. We carried out this evaluation of $s$ for varying values
of $u_{w}$ within the same system and for fixed $u_{w}$ and varying
density. The results of this study, reported in Table \ref{table_s}, clearly
indicate that $s$ does not greatly vary for a broad range of values of the
cavity-border energy $u_{w}$. In fact, at low and moderate values of $u_{w}$
the energy-cavities are isolated and their average size (in $\AA$) grows quite
slowly with $u_{w}$. This is due to the steepness of the hard-core part of the
Lennard-Jones potential.  Above a certain energy $u_{w}$ the cavities become
connected and a steep rise in the average size of the energy-cavities is
observed. This is reflected in the value of $s$.  As shown in Table
\ref{table_s} for $u_{w}=14.19$ Kcal/mol we obtained $s\simeq 4.5$ and
$s\simeq 11$ for two calculations using $d=15$ and $d=100$ respectively. We
obtained a relatively close value $s\simeq 7$ for twice as large an energy limit
$u_{w}=28.38$ Kcal/mol.  However using $u_{w}=165.53$ Kcal/mol the average number of
independent samples increased up to $25$, reflecting the more complex shape
and larger volume of these energy cavities.  In summary, for the optimum range
of values of $u_{w} \sim [10-30]$ Kcal/mol we find $s\simeq [5-10]$
in the case of the Lennard-Jones liquid. 

\begin{table*}
  \begin{tabular}{|c|c|c||c|c|c|c|}
   \hline
 $u_{w}$(Kcal/mol) &  $d$  & $n_0$ &  $\fermac$ & $s$ & $n_{cost}/n_0$ & $\mu$ (Kcal/mol)  \\
    \hline
      28.38 &  20  & $4.2\,10^{5}$  & $1.1\,10^{-3}$   & 7    & 1.5 &  -0.32    \\
      28.38 &  100 & $2.95\,10^{6}$ & $1.08\,10^{-3}$ &  7    & 3.8  &  -0.353  \\
      14.19 &  100 & $4.3\,10^{6}$  & $2.66\,10^{-3}$ &  12  & 1.7  &  -0.335  \\
      14.19 &  15  & $1.0\,10^{7}$  & $2.77\,10^{-3}$ &  5   & 1.1  &  -0.334  \\
      165.53&  200 & $2.76\,10^{5}$ & $1.3\,10^{-5}$  &  25   & 85.8 &  -0.357  \\
    \hline
  \end{tabular}
  \caption{Details of the energy-biased calculations in a Lennard-Jones (LJ)
    liquid at density $\rho=0.0236\AA^{-3}$ and temperature $T=84$K ($\rho=0.92$
    and $T=0.7$ in LJ units). We compare the results for varying 
values of the energy parameter $u_{w}$, samples per cavity $d$ and
varying number $n_0$ of energy probes within the unbiased distribution.  
The cumulative probabilities of the unbiased distribution ($F_{w}=F(u_{w})=\int_{-\infty}^ {u_{w}}f(u)du$) are
    $F_{w}(165.53)=0.0704, F_{w}(28.38)=0.00458, F_{w}(14.19)=0.00122$.  The average 
of the Fermi function in the biased distribution $\fermac$ is
    defined in Eq. (\ref{general}).  The average number of independent samples
    per cavity $s$ is obtained from $s=d/\tau_c$, where the correlation number
    $\tau_c$ is calculated from the correlation between the whole chain of
    data.  The overall number of energy probes in the energy-biased method is
    $n_{cost}=n_0\left(1+d\fstar/\ac\right)$, where $\ac$ is the acceptance
    ratio obtained for the \hr sampler, $\ac\simeq0.17$. The estimation of the
    chemical potential using the standard (unbiased) Bennett method with
    $1.1\times 10^{7}$ energy samples is $\mu=-0.323$ Kcal/mol.}
\label{table_s}
\end{table*}


\begin{thebibliography}{17}
\expandafter\ifx\csname natexlab\endcsname\relax\def\natexlab#1{#1}\fi
\expandafter\ifx\csname bibnamefont\endcsname\relax
  \def\bibnamefont#1{#1}\fi
\expandafter\ifx\csname bibfnamefont\endcsname\relax
  \def\bibfnamefont#1{#1}\fi
\expandafter\ifx\csname citenamefont\endcsname\relax
  \def\citenamefont#1{#1}\fi
\expandafter\ifx\csname url\endcsname\relax
  \def\url#1{\texttt{#1}}\fi
\expandafter\ifx\csname urlprefix\endcsname\relax\def\urlprefix{URL }\fi
\providecommand{\bibinfo}[2]{#2}
\providecommand{\eprint}[2][]{\url{#2}}

\bibitem[{\citenamefont{Lu et~al.}(2003)\citenamefont{Lu, Singh, and
  Kofke}}]{lu03}
\bibinfo{author}{\bibfnamefont{N.}~\bibnamefont{Lu}},
  \bibinfo{author}{\bibfnamefont{J.~K.} \bibnamefont{Singh}}, \bibnamefont{and}
  \bibinfo{author}{\bibfnamefont{D.~A.} \bibnamefont{Kofke}},
  \bibinfo{journal}{J. Chem. Phys.} \textbf{\bibinfo{volume}{118}},
  \bibinfo{pages}{2977} (\bibinfo{year}{2003}).

\bibitem[{\citenamefont{Allen and Tildesley}(1987)}]{allen}
\bibinfo{author}{\bibfnamefont{M.}~\bibnamefont{Allen}} \bibnamefont{and}
  \bibinfo{author}{\bibfnamefont{D.}~\bibnamefont{Tildesley}},
  \emph{\bibinfo{title}{Computer Simulations of Liquids}}
  (\bibinfo{publisher}{Oxford University Press}, \bibinfo{year}{1987}).

\bibitem[{\citenamefont{Frenkel and Smith}(2002)}]{frenkel.book}
\bibinfo{author}{\bibfnamefont{D.}~\bibnamefont{Frenkel}} \bibnamefont{and}
  \bibinfo{author}{\bibfnamefont{B.}~\bibnamefont{Smith}},
  \emph{\bibinfo{title}{Understanding Molecular Simulation: From Algorithms to
  Applications}} (\bibinfo{publisher}{Academic Press, San Diego, 2nd edition},
  \bibinfo{year}{2002}).

\bibitem[{\citenamefont{Kollman}(1993)}]{kollman93}
\bibinfo{author}{\bibfnamefont{P.}~\bibnamefont{Kollman}},
  \bibinfo{journal}{Chem. Rev.} \textbf{\bibinfo{volume}{93}},
  \bibinfo{pages}{2395} (\bibinfo{year}{1993}).

\bibitem[{\citenamefont{Shirts and Pande}(2005)}]{Shirts05}
\bibinfo{author}{\bibfnamefont{M.~R.} \bibnamefont{Shirts}} \bibnamefont{and}
  \bibinfo{author}{\bibfnamefont{V.~S.} \bibnamefont{Pande}},
  \bibinfo{journal}{J. Chem. Phys.} \textbf{\bibinfo{volume}{122}},
  \bibinfo{pages}{144107} (\bibinfo{year}{2005}).

\bibitem[{\citenamefont{Deitrick et~al.}(1989)\citenamefont{Deitrick, Scriven,
  and Davis}}]{dei89}
\bibinfo{author}{\bibfnamefont{G.~L.} \bibnamefont{Deitrick}},
  \bibinfo{author}{\bibfnamefont{L.~E.} \bibnamefont{Scriven}},
  \bibnamefont{and} \bibinfo{author}{\bibfnamefont{H.~T.} \bibnamefont{Davis}},
  \bibinfo{journal}{J. Chem. Phys.} \textbf{\bibinfo{volume}{90}},
  \bibinfo{pages}{2370} (\bibinfo{year}{1989}).

\bibitem[{\citenamefont{Shing and Gubbins}(1982)}]{shing82}
\bibinfo{author}{\bibfnamefont{K.~S.} \bibnamefont{Shing}} \bibnamefont{and}
  \bibinfo{author}{\bibfnamefont{K.~E.} \bibnamefont{Gubbins}},
  \bibinfo{journal}{Mol. Phys.} \textbf{\bibinfo{volume}{46}},
  \bibinfo{pages}{1109} (\bibinfo{year}{1982}).

\bibitem[{\citenamefont{Bennett}(1976)}]{bennett76}
\bibinfo{author}{\bibfnamefont{C.~H.} \bibnamefont{Bennett}},
  \bibinfo{journal}{J. Comput. Phys.} \textbf{\bibinfo{volume}{22}},
  \bibinfo{pages}{245} (\bibinfo{year}{1976}).

\bibitem[{\citenamefont{Jedlovszky and Mezei}(2000)}]{jed00}
\bibinfo{author}{\bibfnamefont{P.}~\bibnamefont{Jedlovszky}} \bibnamefont{and}
  \bibinfo{author}{\bibfnamefont{M.}~\bibnamefont{Mezei}}, \bibinfo{journal}{J.
  Am. Chem. Soc.} \textbf{\bibinfo{volume}{122}}, \bibinfo{pages}{5125}
  (\bibinfo{year}{2000}).

\bibitem[{\citenamefont{Pohorille and Wilson}(1996)}]{poh96}
\bibinfo{author}{\bibfnamefont{A.}~\bibnamefont{Pohorille}} \bibnamefont{and}
  \bibinfo{author}{\bibfnamefont{M.~A.} \bibnamefont{Wilson}},
  \bibinfo{journal}{J. Chem. Phys.} \textbf{\bibinfo{volume}{104}},
  \bibinfo{pages}{3760} (\bibinfo{year}{1996}).

\bibitem[{\citenamefont{Jarynski}(1997)}]{jarz97}
\bibinfo{author}{\bibfnamefont{C.}~\bibnamefont{Jarynski}},
  \bibinfo{journal}{Phys. Rev. Lett.} \textbf{\bibinfo{volume}{78}},
  \bibinfo{pages}{2690} (\bibinfo{year}{1997}).

\bibitem[{\citenamefont{Smith}(1984)}]{Smith84}
\bibinfo{author}{\bibfnamefont{R.~L.} \bibnamefont{Smith}},
  \bibinfo{journal}{Operations Research} \textbf{\bibinfo{volume}{32}},
  \bibinfo{pages}{1296} (\bibinfo{year}{1984}).

\bibitem[{\citenamefont{Goodman}(1960)}]{goodman60}
\bibinfo{author}{\bibfnamefont{L.}~\bibnamefont{Goodman}}, \bibinfo{journal}{J.
  Amer. Stat. Assoc.} \textbf{\bibinfo{volume}{55}}, \bibinfo{pages}{708}
  (\bibinfo{year}{1960}).

\bibitem[{\citenamefont{Kremer and Grest}(1990)}]{grest1}
\bibinfo{author}{\bibfnamefont{K.}~\bibnamefont{Kremer}} \bibnamefont{and}
  \bibinfo{author}{\bibfnamefont{G.}~\bibnamefont{Grest}}, \bibinfo{journal}{J.
  Chem. Phys.} \textbf{\bibinfo{volume}{92}}, \bibinfo{pages}{5057}
  (\bibinfo{year}{1990}).

\bibitem[{\citenamefont{Delgado-Buscalioni and Coveney}(2003)}]{usher}
\bibinfo{author}{\bibfnamefont{R.}~\bibnamefont{Delgado-Buscalioni}}
  \bibnamefont{and} \bibinfo{author}{\bibfnamefont{P.~V.}
  \bibnamefont{Coveney}}, \bibinfo{journal}{J. Chem. Phys.}
  \textbf{\bibinfo{volume}{119}}, \bibinfo{pages}{978} (\bibinfo{year}{2003}).

\bibitem[{\citenamefont{{De Fabritiis} et~al.}(2004)\citenamefont{{De
  Fabritiis}, Delgado-Buscalioni, and Coveney}}]{water_usher}
\bibinfo{author}{\bibfnamefont{G.}~\bibnamefont{{De Fabritiis}}},
  \bibinfo{author}{\bibfnamefont{R.}~\bibnamefont{Delgado-Buscalioni}},
  \bibnamefont{and} \bibinfo{author}{\bibfnamefont{P.~V.}
  \bibnamefont{Coveney}}, \bibinfo{journal}{J. Chem. Phys.}
  \textbf{\bibinfo{volume}{121}}, \bibinfo{pages}{12139}
  (\bibinfo{year}{2004}).

\bibitem[{\citenamefont{Liu}(2001)}]{liu01}
\bibinfo{author}{\bibfnamefont{J.~S.} \bibnamefont{Liu}},
  \emph{\bibinfo{title}{Monte Carlo Strategies in Scientific Computing}}
  (\bibinfo{publisher}{New York: Springer-Verlag}, \bibinfo{year}{2001}).

\end{thebibliography}
\end{document}